\documentclass[prb,aps,twocolumn,showpacs,preprintnumbers, superscriptaddress, amsmath,amssymb,10pt,letterpaper]{revtex4-1}
\usepackage{comment}
\usepackage{xcolor} 
\usepackage{soul}

\usepackage{graphicx}
\usepackage{dcolumn}
\usepackage{bm}
\usepackage{dsfont}
\usepackage{textcomp} 

\def\bea{\begin{equation}\begin{aligned}}
\def\eea{\end{aligned}\end{equation}}

\begin{document}
\large
\title{Spin Accumulation and Longitudinal Spin Diffusion of Magnets}
\author{Wayne M. Saslow}
\email{wsaslow@tamu.edu}
\affiliation{ Texas A\&M University, College Station, Texas, 77843, U.S.A. }
\author{Chen Sun}
\email{chensun@hnu.edu.cn} 
\affiliation{ School of Physics and Electronics, Hunan University, Changsha 410082, China}
\author{Shenglong Xu}
\email{slxu@tamu.edu}
\affiliation{ Texas A\&M University, College Station, Texas, 77843, U.S.A. }
\begin{abstract}

We extend to the  longitudinal component of the magnetization  the spintronics idea that a magnet near equilibrium can be described by two magnetic variables.  One is the usual magnetization $\vec{M}$. 
The other is the non-equilibrium quantity $\vec{m}$, called the spin accumulation, by which the non-equilibrium spin current can be transported.  $\vec{M}$ represents a correlated distribution of a very large number of degrees of freedom, as expressed in some equilibrium distribution function for the excitations; we therefore forbid $\vec{M}$ to diffuse, but we permit $\vec{M}$ to decay.  On the other hand, we  permit $\vec{m}$, due to spin excitations, to both diffuse and decay.  For this physical picture, diffusion from a given region occurs by decay of $\vec{M}$ to $\vec{m}$, then by diffusion of $\vec{m}$, and finally by decay of $\vec{m}$ to $\vec{M}$ in another region.  This somewhat slows down the diffusion process. 
Restricting ourselves to the longitudinal variables $M$ and $m$ with equilibrium properties $M_{eq}=M_{0}+\chi_{M\parallel}H$ and $m_{eq}=0$, we argue that the effective energy density must include a new, thermodynamically required exchange constant $\lambda_{M}=-1/\chi_{M\parallel}$.  We then develop the appropriate macroscopic equations by applying Onsager's irreversible thermodynamics,
and use the resulting equations to study the space and time response.  At fixed real frequency $\omega$ there is, as usual, a single pair of complex wavevectors $\pm k$ but with an unusual dependence on $\omega$.  At fixed real wavevector, there are two decay constants, as opposed to one in the usual case. Extending the idea that non-equilibrium diffusion in other ordered systems involves a non-equilibrium quantity, this work suggests that in a superconductor the order parameter $\Delta$ can decay but not diffuse, but a non-equilibrium gap-like $\delta$, due to pair excitations, can both decay and diffuse. 

\end{abstract}

\date{\today}

\maketitle






\section{Introduction}
\label{sect:IntroLongOnly}
The technologically important field of spintronics uses spin currents to make magnets receive (read) or send (write) information.  Associated with this is the major theoretical advancement that, when out of equilibrium, a magnet has a second magnetic variable, called the spin accumulation.  The term {\it accumulation of spin} appeared in the 1970 work of Dyakonov and Perel on what are now known as the spin Hall effect (an electric current produces a magnetization in a non-magnetic spin-active material) and the inverse spin Hall effect.\cite{DyakonovPerel1,DyakonovPerel2}

Transverse spin currents (relative to the magnetization $\vec{M}$) were implicit in the early work of Monod {\it et al},\cite{Monod72} and were explicit in the 1979 work of Silsbee {\it et al},\cite{SilsbeeJanossyMonod79} both of which involved magnetization deviations transverse to the equilibrium magnetization.  The latter work invoked an exchange interaction between two posited types of magnetic electrons ($s$ and $d$ were spin-polarized but only $s$ could conduct -- i.e., diffuse).\cite{Hasegawa59}  Longitudinal spin currents were studied theoretically by Johnson and Silsbee.\cite{JohnsonSilsbee87,JohnsonSilsbee88}  The 1993 theory of Valet and Fert\cite{ValetFert93} considered longitudinal spin currents and introduced the term {\it spin accumulation}.\cite{ValetFertspinacc}

In 2002 Zhang, Levy and Fert employed an $s$-$d$ model with exchange to study (transverse) spin transfer torque.\cite{ZhangLevyFert02} In it the dominant magnetization was the non-diffusing $\vec{M}_{d}$ (with a subscript suggesting the real-space core $d$ electrons), and the spin current was due only to the magnetization $\vec{m}$ of the conduction $s$ electrons.  It was followed in 2004 by a kinetic theory for an itinerant magnet that replaced $\vec{M}_{d}$ by the usual magnetization symbol $\vec{M}$; and the excitations were described by a distribution function in momentum space.  The words ``spin accumulation'' were employed, but not given a symbol or identification.\cite{ZhangLevyFertspinacc}   The spin current was explicitly given in terms of the distribution function.

$\vec{M}$ can be taken to be due to the momentum-space ``core'' electrons within the majority and minority Fermi seas, which in an $s$-$d$ model would include both the $s$ and $d$ bands, and $\vec{m}$ can be taken to be due to momentum-space excitations at the Fermi surfaces. Both Refs.~\onlinecite{ZhangLevyFert02,ZLZA04} emphasized the transverse components of $\vec{M}$ and $\vec{m}$, which are the magnetic variables used in the area of spintronics.  $\vec{M}$ is described by the classic theory of Landau and Lifshitz.\cite{LLMagnetics35}

Applying irreversible thermodynamics we previously examined the full equations of motion for this model.\cite{Saslow17} Unlike the present work, that work assumed a single longitudinal type of magnetization.
More recently we have studied the coupled transverse modes of $\vec{M}$ and $\vec{m}$, as can be generated in an ac spin transfer torque or spin pumping experiment.\cite{SunSaslow19}

Fig.~1 of that work provides a physical picture, for a two-band conducting magnet, of the two different ways (tipping of $\vec{M}$ and excitations that cause spin accumulation $\vec{m}$) to obtain a net transverse magnetization, but the idea also applies to insulating magnets. For this $\vec{M}$-$\vec{m}$ model we take the net magnetization $\vec{\cal M}$ to be given by
\begin{equation}
\vec{\cal M}=\vec{M}+\vec{m}.
\label{calM}
\end{equation} 
It is well-known in electrical and thermal conduction that only the non-equilibrium part of the statistical distribution function gives rise to the diffusive processes leading to the electric current and the heat current. An analogous argument is that for a magnet it is the non-equilibrium part of the statistical distribution function that gives rise to both $\vec{m}$ and the diffusive processes leading to the spin current.

We have therefore developed the idea that because the equilibrium magnetization $M$ represents a distribution of a macroscopically large number of degrees of freedom, it cannot diffuse; it can only grow or decay.\cite{size}  Thus only the spin accumulation $m$, due to the non-equilibrium part of the statistical distribution function, is permitted to diffuse.
On the other hand, because $\vec{m}$ must have a source, such as a local fluctuation of $\vec{M}$, by reciprocity $\vec{m}$ must also be able to provide a source for $\vec{M}$, and thus $\vec{m}$ can both decay and diffuse.

\begin{figure}[!htb]
\scalebox{0.38}[0.38]{\includegraphics{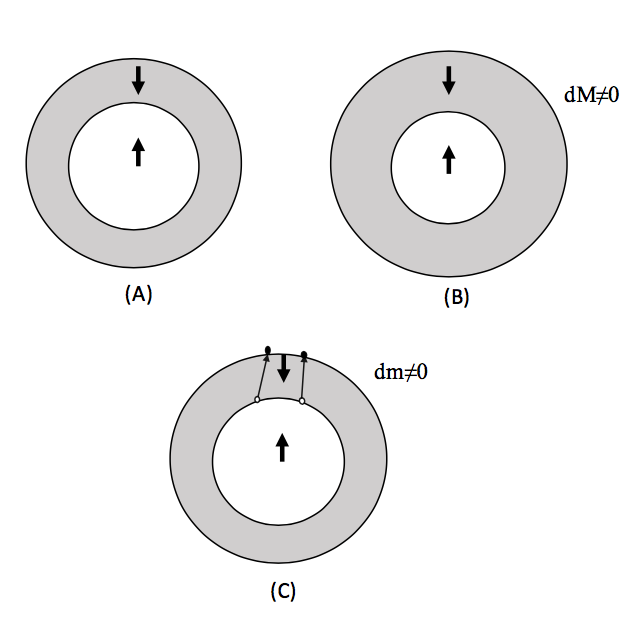}}
\hspace{-0.5cm}\vspace{-0.8cm}
\caption{  Two ways to change net magnetization.  (A) Fermi sea for $dM=0$, $dm=0$.  (B) Fermi sea for $dM\ne0$, $dm=0$.   This is obtained by enlarging the down spin Fermi sea at the expense of the up spin Fermi sea, their net volume being unchanged.  (C). Fermi sea for $dM=0$, $dm\ne0$.   This is obtained by transitions from the up spin Fermi sea to the down spin Fermi sea.  The energy difference is small because the difference in exchange energy is nearly compensated by the difference in kinetic energy. } 
\vspace{-3mm}
\label{fig:LongitudinalTwoWays}
\end{figure}

Fig.~\eqref{fig:LongitudinalTwoWays} illustrates, for a two-band magnet, the two ways by which a magnet can obtain a net longitudinal magnetization.

As a consequence a non-uniform magnetic system can be thought to relax in the following way.  Imagine that everywhere $\vec{M}$ takes on a uniform value, except for a small region A where it takes a larger uniform value.  For equilibration between the regions to occur, in region A $\vec{M}$ must decay to $\vec{m}$, which in turn can diffuse out of region A.  After $\vec{m}$ has diffused to region B it can decay to $\vec{M}$, thus transferring the excess in A to region B.  The net diffusion, however, must include all fluctuations.  
Sect.~III illustrates the behavior of both $M$ and $m$ for initial conditions where $m=0$ everywhere and where is $M$ is initially in equilibrium with a specific field $H$ that suddenly is removed. 

\hspace{-2cm}\vspace{-0.8cm}

In what follows we establish the near-equilibrium thermodynamics of this system (Sect.~II), and in Sect.~III we discuss the nature of the out-of-equilibrium statistical effective field that couples $m$ to $M-M_{0}$.  It is distinct from the dynamical exchange field $\lambda\vec{M}$ that causes transverse $\vec{M}$ to precess around $\vec{M}$. We then apply Onsager's irreversible thermodynamics to obtain the equations of motion for $M$ and $m$ (Sect.~IV), where we study their coupled response to oscillating space variations $e^{ikx}$ for real $k$ and to oscillating time variations $e^{-i\omega t}$ for real $\omega$.  We then discuss the boundary conditions that must be satisfied (Sect.~V),  provide a brief summary (Sect.~VI), and point out the implications for diffusion in other ordered systems, such as superconductors (Sect.~VII).

\section{Near Equilibrium Thermodynamics: Longitudinal Variables}
\label{sect:NearEquilLong}

Consider a collinear magnet in a field $\vec{H}$ with remanence $M_{0}$ and magnetic susceptibility $\chi_{M}$, and an effective exchange interaction between $M$ and $m$.  Then any macroscopic effective energy, when minimized, must give the equilibrium values
\begin{equation}
M_{eq}=M_{0}+\chi_{M}H, \quad m_{eq}=0.
\label{equilibrium}
\end{equation}
Further, the minimization conditions for both $M$ and $m$ must be consistent with one another.  This constrains the effective energy, and leads to an exchange term that has not been considered previously.

We take $\mu_{0}$ to be the permeability of free space, and employ SI units, where fields and magnetization are in units of A/m.  We then take the system to have an effective energy density whose dependence on $M$ and $m$ is given by
\begin{eqnarray}
\varepsilon&=&-\mu_{0}({M}+{m}){H}
+\frac{\mu_{0}}{2} \Big[ \frac{(M-M_{0})^{2}}{\chi_{M\parallel}}+\frac{m^{2}}{\chi_{m\parallel}}\Big]\cr
&&-\mu_{0}\lambda_{M}(M-M_{0}){m}.
\label{eps}
\end{eqnarray}
Here $\chi_{M\parallel}$ and $\chi_{m\parallel}$ are dimensionless susceptibilities associated with the parallel direction (for stability they must be non-negative), $M_{0}$ is the spontaneous magnetization, and $\lambda_{M}$ is a dimensionless mean-field coefficient whose value is determined by the equilibrium condition $m_{eq}=0$.  With the exception of the terms in $m$, this form is well-known to give $M=M_{0}+\chi_{M\parallel}H$ in equilibrium.  For the remainder of this work we drop the subscript $\parallel$, although we note that magnets can respond in more than one direction.


We define effective fields $H^{*}$ and $h^{*}$
\begin{eqnarray}
H^{*}&\equiv&-\frac{1}{\mu_{0}}\frac{\partial\varepsilon}{\partial M}=H+\lambda_{M} m-\frac{M-M_{0}}{\chi_{M}}\cr
&\equiv&-\frac{\delta M}{\chi_{M}},
\label{M_L}\\
h^{*}&\equiv&-\frac{1}{\mu_{0}}\frac{\partial\varepsilon}{\partial m}=H+\lambda_{M} (M-M_{0})-\frac{m}{\chi_{m}}\cr
&\equiv&-\frac{\delta m}{\chi_{m}}. \quad
\label{m_L}
\end{eqnarray}
In equilibrium we want the parameters of the theory to ensure that $H^{*}=0$ and $h^{*}=0$.


Eq.~\eqref{M_L} satisfies the local equilibrium condition \eqref{equilibrium} for any $\lambda_{M}$.  However, Eq.~\eqref{m_L} does {\it not} satisfy  and $h^{*}=0$ for any $\lambda_{M}$.  We ensure that Eq.~\eqref{m_L} also satisfies $h^{*}=0$ by requiring that 
\begin{equation}
\lambda_{M}=-\frac{1}{\chi_{M}}.
\label{lambdaL}
\end{equation}
This value $\lambda_{M}=-\frac{1}{\chi_{M}}$ is thermodynamically required.  
It may be thought of as specifying a mean field acting on $m$ that, for the equilibrium value of $M$, ensures that $m=0$. 

We now define
\begin{equation}
\Delta m\equiv m-m_{eq}=m, \quad \Delta M\equiv M-M_{eq}.
\label{dMdm}
\end{equation}
When the differences are differentials we may replace $(\Delta M, \Delta m)$ by $(dM, dm)$.

With the definition
\begin{equation}
\xi\equiv\frac{\chi_{m}}{\chi_{M}}
\label{xi}
\end{equation}
we have
\begin{eqnarray}
\delta M&\equiv& -\chi_{M}H^{*}\cr
&=&(M-M_{0}-\chi_{M}H)+m\cr
&\equiv&\Delta M+\Delta m,
\label{delM_L}\\
\delta m&\equiv& -\chi_{m}h^{*}\cr
&=&m+\frac{\chi_{m}}{\chi_{M}}(M-M_{0}-\chi_{M}H)\cr
&=&\Delta m+\xi \Delta M.
\label{delm_L}
\end{eqnarray}
On employing \eqref{lambdaL} in \eqref{eps}, and studying small fluctuations of $m$ and $M$, we find that thermodynamic stability requires that  $0\le\xi\le1$, or $\chi_{m}\le\chi_{M}$.

\section{On the longitudinal effective fields $H^{*}$ and $h^{*}$}
There is, of course, a longitudinal exchange field $H_{ex}=\lambda_{M} M$ that acts on $m$.  In the dynamical equation for $\vec{M}$ it is responsible for precessional motion of the transverse spin accumulation.  \\
\indent{}However, the effective energy density $\varepsilon$ has been constructed, with thermodynamic equilibrium in mind -- which basically is a matter of statics -- to give net effective fields $H^{*}$ and $h^{*}$ that yield $m_{eq}=0$ and $M_{eq}=M_{0}+\chi_{M}H$. This is done by having a statistical field that is non-zero only out of equilibrium, and gives the correct $m$ and $M$ in equilibrium.  \\
\indent{}We emphasize that this statistical field, which is zero in equilibrium, is distinct from the dynamical exchange field that is present even in equilibrium.  This exchange field $\lambda\vec{M}$, if a transverse $\vec{m}$ develops, can cause $\vec{m}$ to precess around $\vec{M}$.   For a ferromagnet with up and down Fermi surfaces we expect that $\lambda$ is proportional to the sum over the Fermi surfaces of the product of a Fermi liquid constant and a density of states. The exchange field of Ref.~\onlinecite{Saslow17,SunSaslow19} is based on the thermally averaged exchange interaction, as computed, for example, in Fermi liquid theory, and given in Ref.~\onlinecite{ZhangLevyFert02}.  The present exchange field is obtained from the free energy density $f=F/V$, where for a uniform system $F=-k_{B}T\ln[{\rm Tr}(e^{-E/k_{B}T})]$ and $V$ is the volume.\\

\section{Longitudinal Dynamics}
\label{sect:IrTh}
We now derive the equations of motion for the longitudinal magnetic response of $M$ and $m$.  We employ Onsager's irreversible thermodynamics.\cite{Onsager1,Onsager2}

In the thermodynamic relation for the differential of the energy density $d\varepsilon$ we employ the previously defined effective fields $H^{*}$ and $h^{*}$ to write
\begin{equation}
d\varepsilon=Tds - \mu_{0}H^{*}dM - \mu_{0}h^{*}dm.
\label{thermoMm}
\end{equation}
Since in equilibrium $H^{*}=0$ and $h^{*}=0$, $\varepsilon$ is minimized on varying $M$ and $m$.

As already noted, because $M$ characterizes a distribution function for a macroscopic number of excitations, we consider that it cannot diffuse; a structure with $10^{23}$ variables is not expected to diffuse.  That is not true of the excitations, which yield $m$.  With this in mind, we now write down the ``conservation laws'' for energy density $\varepsilon$, entropy density $s$, $M$, and $m$.

With unknown source terms $R$ and flux terms $j$ as appropriate, we take 
\begin{eqnarray}
\partial_{t}\varepsilon + \partial_{i}j^{\varepsilon}_{i}=0,
\label{depsdt}\\
\partial_{t}{s} + \partial_{i}j^{s}_{i}=R_{s}\ge0,
\label{dsdt}\\
\partial_{t}{M} =R_{M},
\label{dMdt}\\
\partial_{t}{m} + \partial_{i}j^{m}_{i} =R_{m},
\label{dmdt}
\end{eqnarray}
where $j^{\varepsilon}_{i}$, $j^{s}_{i}$ and $j^{m}_{i}$ are the energy flux, entropy flux and spin accumulation flux respectively, and $R_{s}$, $R_{M}$, $R_{m}$ are the production rates of entropy, magnetization and spin accumulation, respectively. Above we used the principle that the rate of entropy production is non-negative, or $R_{s}\ge0$.

Using these equations we find that the time-derivative of \eqref{thermoMm} can be rewritten as
\begin{eqnarray}
0\le TR_{s}&=& T\partial_{t}s+T\partial_{i}j^{s}_{i}\cr
&=&\partial_{t}\varepsilon+\mu_{0}H^{*}\partial_{t}M + \mu_{0}h^{*}\partial_{t}m+T\partial_{i}j^{s}_{i}\cr
&=&-\partial_{i}(j^{\varepsilon}_{i}-Tj^{s}_{i}+\mu_{0}h^{*}j^{m}_{i})\cr
&& -j^{s}_{i}\partial_{i}T+\mu_{0}j^{m}_{i}\partial_{i}h^{*}\cr
&& +\mu_{0}R_{M}H^{*}+\mu_{0}R_{m}h^{*}.
\label{R}
\end{eqnarray}

Observing the space and time properties of the $R$'s and $j$'s, we obtain linear relations between the unknown fluxes and the unknown sources:
\begin{eqnarray}
j^{s}_{i}&=&-\frac{\kappa}{T}\partial_{i}T,
\label{js}\\
j^{m}_{i}&=&C_{m}\partial_{i}h^{*}\equiv -D\partial_{i}(\delta m), \quad
\label{jm}\\
R_{M}&=&L_{MM}H^{*} + L_{Mm}h^{*},
\label{R_M}\\
R_{m}&=&L_{mm}h^{*} + L_{mM}H^{*},
\label{R_m}
\end{eqnarray}
where $\kappa$ is the thermal conductivity, $C_m=-D/\chi_{m}$, $D$ is the diffusion coefficient, and $L_{MM}$, $L_{Mm}$, $L_{mm}$ and $L_{mM}$ are Onsager constants relating $R_{M}$ and $R_{m}$ to $H^{*}$ and $h^{*}$. There is an Onsager relation that
\begin{equation}
L_{mM}=L_{Mm}\equiv -L,
\label{OnsagerL}
\end{equation}
so there are only three independent constants associated with the sources.  (The minus sign is expected for cross-decay.) We also neglect the thermomagnetic Onsager constants that relate entropy flux $j^{s}_{i}$ to $\partial_{i}h^{*}$, and spin accumulation flux $j^{m}_{i}$ to $\partial_{i}T$.

Using Eqs.~\eqref{M_L} and \eqref{m_L}, Eqs.~\eqref{R_M} and \eqref{R_m} can be written as:
\begin{eqnarray}
R_{M}&=&-\delta M\frac{L_{MM}}{\chi_{M}}-\delta m\frac{L_{Mm}}{\chi_{m}},
\label{RM3}\\
R_{m}&=&-\delta m\frac{L_{mm}}{\chi_{m}}  -\delta M\frac{L_{mM}}{\chi_{M}}.
\label{Rm3}
\end{eqnarray}

$R_{M}$ and $R_{m}$ may  be rewritten in a more transparent way by introducing four (related) relaxation times.
Thus we may rewrite $R_{M}$ and $R_{m}$ as
\begin{eqnarray}
R_{M}\approx -\delta M( \frac{1}{\tau_{Mm}}+\frac{1}{\tau_{ML}} )+\frac{\delta m}{\tau_{mM}},
\label{RM2}\\
R_{m}\approx -\delta m(\frac{1}{\tau_{mM}}+\frac{1}{\tau_{mL}})+\frac{\delta M}{\tau_{Mm}}.
\label{Rm2}
\end{eqnarray}

Comparing Eqs.~\eqref{RM2} and \eqref{Rm2} with Eqs.~\eqref{RM3} and \eqref{Rm3}, we have
\begin{eqnarray}
\frac{1}{\tau_{Mm}}+\frac{1}{\tau_{ML}} &=&\frac{L_{MM}}{\chi_{M}},
\quad - \frac{L_{Mm}}{\chi_{m}}=\frac{1}{\tau_{mM}},\label{OnsagerM}\\
\frac{1}{\tau_{mM}}+\frac{1}{\tau_{mL}}&=&\frac{L_{mm}}{\chi_{m}},
\quad - \frac{L_{mM}}{\chi_{M}}=\frac{1}{\tau_{Mm}}\label{Onsagerm}.
\end{eqnarray}

The Onsager constant of \eqref{OnsagerL} takes the form
\begin{align}
L\equiv\frac{ \chi_{m} }{ \tau_{mM} }=\frac{ \chi_{M} }{ \tau_{Mm} }.
\label{L}
\end{align}

\subsection{Linearized Equations of Motion}
\label{sect:EqMotLong}


By \eqref{dMdt} and \eqref{RM2} we have
\begin{equation}
\frac{\partial M}{\partial t}=R_{M}= -\delta M( \frac{1}{\tau_{Mm}}+\frac{1}{\tau_{ML}} )+\frac{\delta m}{\tau_{mM}}.
\label{dMdt2}
\end{equation}

By \eqref{dmdt}, \eqref{jm}, and \eqref{Rm2} we have
\begin{eqnarray}
\frac{\partial m}{\partial t}&-&D\nabla^{2} \delta m=R_{m}\cr
&=& -\delta m(\frac{1}{\tau_{mM}}+\frac{1}{\tau_{mL}})+\frac{\delta M}{\tau_{Mm}}. \quad
\label{dmdt2}
\end{eqnarray}

We now introduce the difference in inverse susceptibilities $\tilde{\chi}^{-1}$
\begin{align}
    \tilde{\chi}^{-1}\equiv(\chi_m^{-1}-\chi_M^{-1}),
    \label{defs1}
\end{align}
where by $\xi=\chi_{m}/\chi_{M}\le1$ we have $\tilde{\chi}\ge0$.

These equations have a natural exchange-driven rate $r$ (between $M$ and $m$) and a natural wavevector $k_{M}$ given by
\begin{equation}
    r\equiv \frac{L}{\tilde{\chi}}
    =\frac{1}{\tau_{mM}}-\frac{1}{\tau_{Mm}}, \qquad k_{M}^{2}\equiv
\frac{r}{D}.
    \label{kL2}
\end{equation}
Since $L\ge 0$ and $\tilde{\chi}>0$, we have $r\ge0 $.  Thus $m$ decays to $M$ more rapidly than $M$ decays to $m$ ($\tau_{mM}\le \tau_{Mm}$).  Such decay is expected to be due to the microscopic exchange interaction, and is likely to be the fastest of the decay times in the system.

In the next sections we will use \eqref{dMdt2} and \eqref{dmdt2} to obtain the time-response of an otherwise uniform system subject to a spatially-oscillating disturbance, and the spatial-response of an otherwise uniform system subject to a time-oscillating disturbance.

\subsection{General Temporal Response}
Consider small deviations from equilibrium $dM$and $dm$.  Then \eqref{delM_L} and \eqref{delm_L} give $\delta M=dM+dm$ and $\delta m=dm+\xi dM$.  On switching to the variables $dM$ and $dm$, the equations of motion \eqref{dMdt2} and \eqref{dmdt2} can be written as
\begin{eqnarray}
(\frac{\partial}{\partial t}&+&\frac{1}{\tau_{ML}})dM\cr
&=&(-\frac{1}{\tau_{ML}}+r)dm,
\label{dMdt3}\\
(\frac{\partial}{\partial t}&+&\frac{1}{\tau_{mL}}+r - D\nabla^{2})dm\cr
&=&-\xi(\frac{1}{\tau_{mL}}-D\nabla^{2})dM.
\label{dmdt3}
\end{eqnarray}

\begin{figure}[!htb]
(a)\scalebox{0.4}[0.4]{\includegraphics{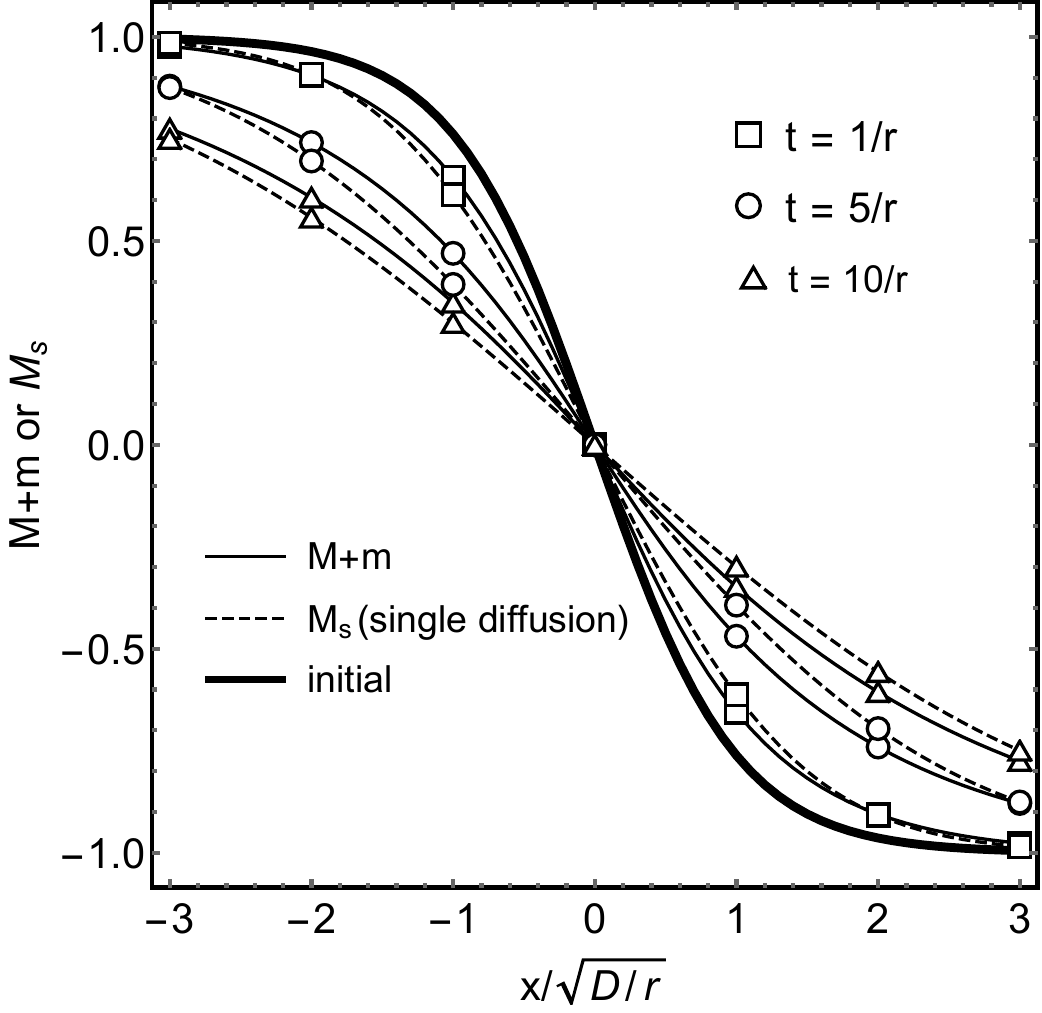}}\\
(b)\scalebox{0.4}[0.4]{\includegraphics{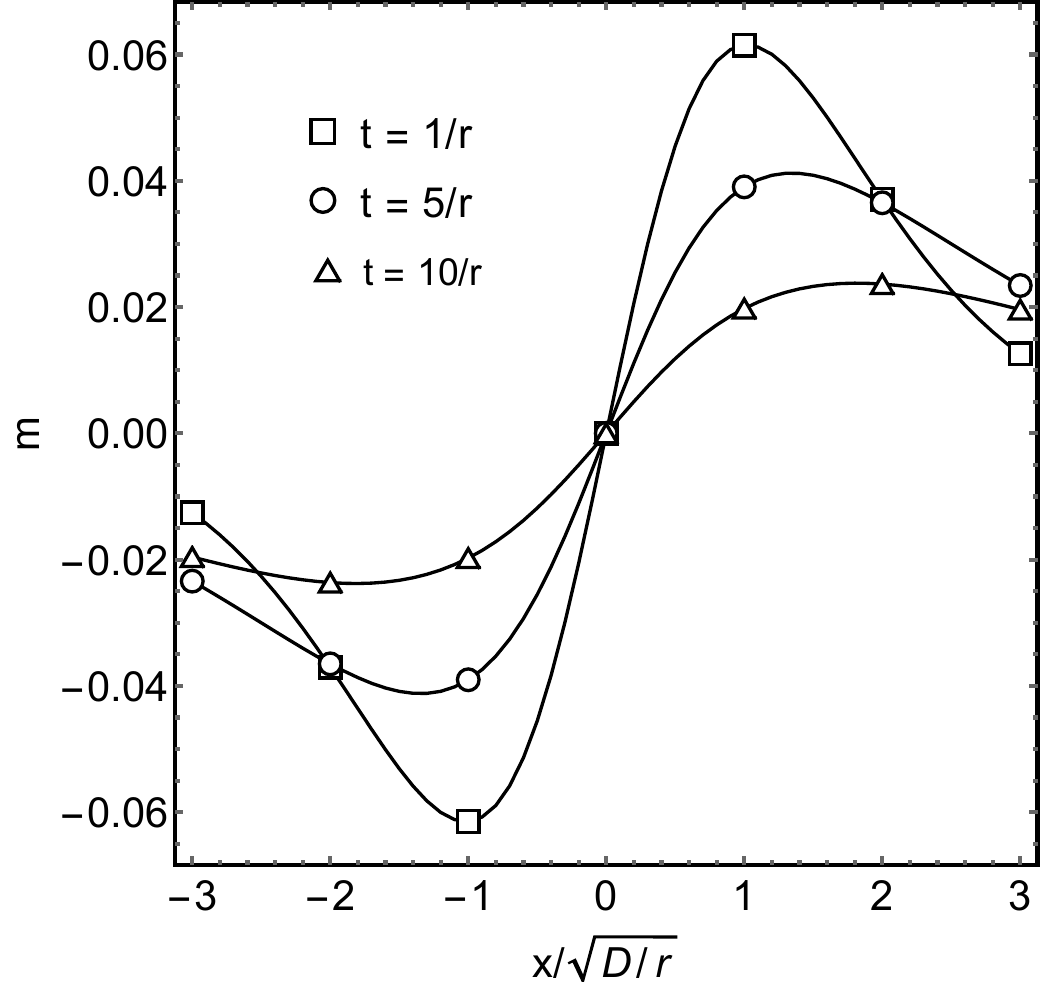}}
\caption{ Time evolution of $M$ and $m$ in the $M$-$m$ model, and in the single diffusion model, both for   $\xi\equiv\chi_{m}/\chi_{M}=0.3$, so $D_{s}=0.3 D$.  The initial conditions are $m=0$ for all $x$, and $M=-\tanh(x)$, initial field $H_{0}\propto M$ that is suddenly removed.  (a) $M(x)+m(x)$ and $M_s(x)$ at the three times $1/r$, $5/r$, and $10/r$. The solid lines are $M(x)+m(x)$, the dashed lines are $M_s$ from the single diffusion model. The thick line is the initial value $M(x)+m(x)=M_s(x)=-\tanh(x)$. (b) $m(x)$ for $M$-$m$ model at the same three times. Note that $D$ is scaled out. }
\label{fig: Mandm3times}
\end{figure}

To illustrate the implications of these equations, we neglect decay to the lattice ($\tau_{ML}, \tau_{mL}\rightarrow\infty$) and consider that $\xi=\chi_{m}/\chi_{M}=0.3$.
We employ initial conditions $m=0$ for all $x$, and $M=-\tanh(x)$ proportional to an initial field $H_{0}\propto M$.  We then suddenly set $H=0$.\\
\indent{}For the three times $1/r$, $5/r$, and $10/r$, Fig.~\ref{fig: Mandm3times}(a) gives profiles of $M(x)+m(x)$ as solid lines. $D$ is scaled out.  It also gives profiles of $M_{s}$, where $s$ stands for ``single diffusion''.  We take $M_{s}$ to satisfy the diffusion equation $\partial M_{s}/\partial t= D_{s}\nabla^{2}M_{s}$ with, for purposes of comparison, $D_{s}=\xi D$. Clearly the $M$-$m$ theory gives results similar to, but distinct from, those for the $M_{s}$ theory; measurement of such a profile can thus distinguish between the two theories.   \\
\indent{}For the same three times, Fig.~\ref{fig: Mandm3times}(b) gives profiles of $m(x)$.  The decay is slower in the present model, with both $M$ and $m$, than in the simple diffusion  model for $M_{s}$ alone.  This slower decay occurs because, in this $M$-$m$ model, $M$ cannot diffuse, but must first decay to $m$, which only then can diffuse.  Although not shown in the figures, the faster the decay rate $r$ from $M$ to $m$, the more quickly the system equilibrates.\\
\indent{} We also considered the initial condition $M=\operatorname{sech}(x)$, with an initial field $H_{0}\propto M$ that is suddenly removed;  here $M$ initially is larger in a small region.  In addition, we calculated the spin currents in the two models. They all show that the decay is slower in the $M$-$m$ model than in the single diffusion model.

\subsection{Temporal Response to Wavevector $k$}
Eqs.~\eqref{dMdt3} and \eqref{dmdt3} can be put in the form of the eigen-equations
\bea
\partial_t \begin{pmatrix} dM \\ dm\end{pmatrix} = \Gamma \begin{pmatrix} dM \\ dm\end{pmatrix}.
\eea
We now consider the temporal response to a disturbance at wavevector $\vec{k}$, where the latter is real.
To do so we introduce an inverse decay rate $\gamma$ (not the gyromagnetic ratio).
Assuming the space and time variation of $dm$ and $dM$ are given by  $\exp(-\gamma t + i\vec{k}\cdot\vec{r})$, the matrix $\Gamma$ is given by
\bea
&\Gamma =
\begin{pmatrix}
-\tau_{ML}^{-1} &  -\tau_{ML}^{-1} + r\\
-\xi (\tau_{mL}^{-1} + Dk^2) & -r -(\tau_{mL}^{-1}+Dk^2)
\end{pmatrix}.
\eea
When $\tau_{ML}\rightarrow \infty$ and $\tau_{mL}\rightarrow \infty$ (i.e. the magnetizations do not decay to the lattice), for $k=0$ we have  $\partial_t(M+m)=0$, a result of conservation of magnetization.
If the wavevector $\vec{k}$ is known, then the unknown decay rate $\gamma(k)$ as a function of wavevector $k$ can be obtained by diagonalizing the matrix $\Gamma$.

\subsubsection{Slow Lattice Decay}
We first work in the limit that $\tau_{ML}, \tau_{mL}\rightarrow\infty$.  This is appropriate to spin-aligned nuclear systems, with small magnetic moments and therefore weak interactions with the lattice. 
Then the eigenrates are
\bea
\gamma^{\pm} = \frac{r}{2}\Bigg[( 1+\frac{ k^{2} }{ k_{M}^{2} } )
\pm \sqrt{ ( 1+\frac{ k^{2} }{ k_{M}^{2} } )^{2}-4\frac{ k^{2}\xi }{ k_{M}^{2} }}\Bigg].
\label{gammaNoL}
\eea

For each mode, by substituting each decay rate in either of the mode equations we can determine that mode's ratio of $dm$ to $dM$. The initial conditions on $dm$ and $dM$ then determine the amplitude of each mode.

Let $\gamma^+$ and $\gamma^-$ be the decay rates for the fast and slow decay modes.  In the long wavelength $k\rightarrow 0$ limit, we have
\bea
\gamma^+ \rightarrow r, \quad \gamma^- \rightarrow (D\xi) k^2.
\label{gamma1}
\eea
The fast decay mode $\gamma^+$ has $dm^{+}\approx -dM^{+}$, so $\delta M\approx 0$.  The slow decay mode $\gamma^-$ is a diffusion mode with $dm^{-}\approx -\xi (k^{2}/k_{M}^{2})dM^{-}$, so the effective diffusion constant is $D\xi=D(\chi_{m}/\chi_{M})$.

\begin{figure}[!htb]
\scalebox{0.35}[0.35]{\includegraphics{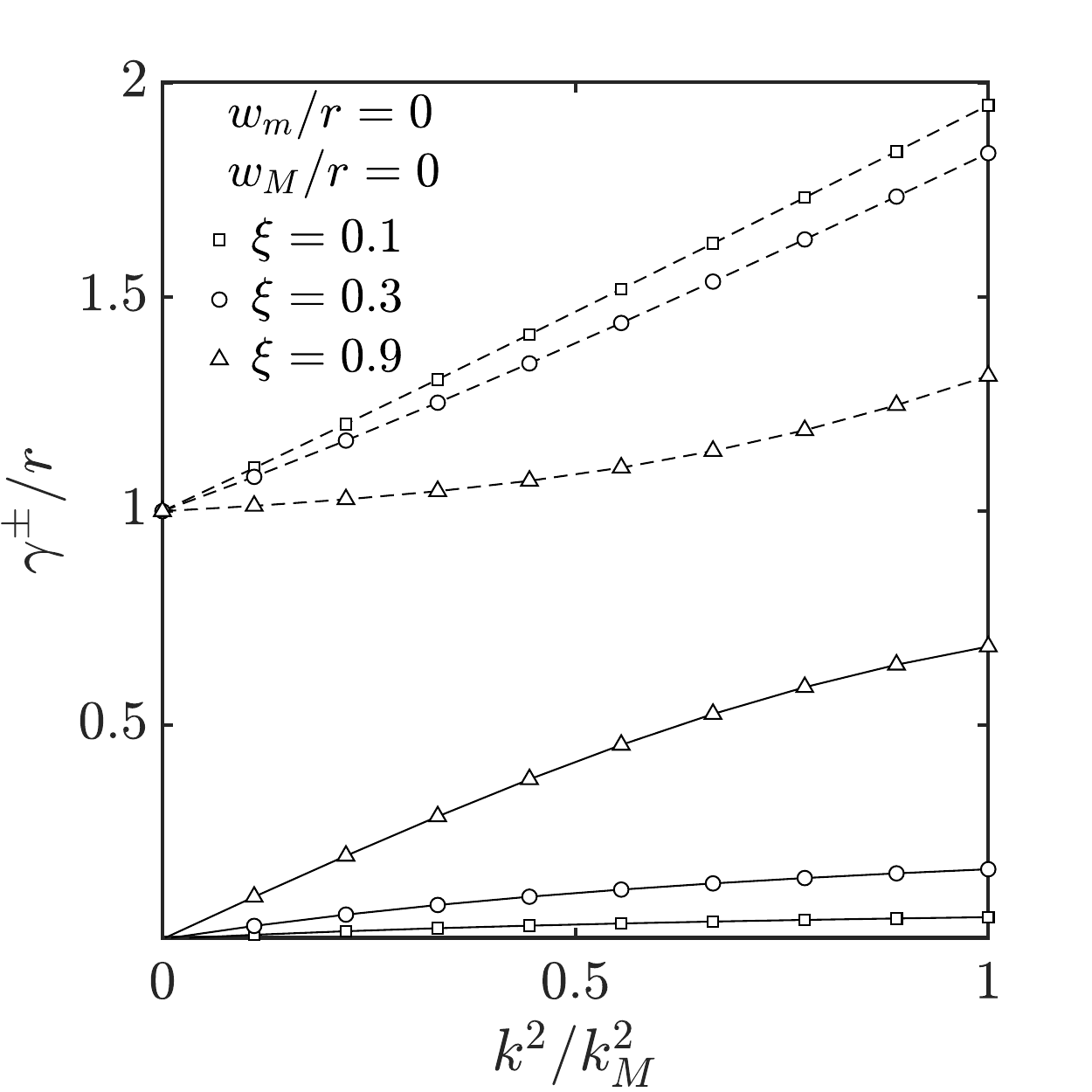}}
\caption{Eigenrates $\gamma$ relative to exchange-driven rate $r$, versus wavevector ratio $(k/k_{M})^{2}$, for three susceptibility ratios $\xi$.   Lattice decay rates $w_{m}$ and $w_{M}$ are neglected.  For $r$ and $\xi$ see \eqref{xi}; for wavevector $k_{M}$ see \eqref{kL2}. }
\label{fig: no_latticedecay}
\end{figure}

Fig.~\ref{fig: no_latticedecay} presents $\gamma/r$ for no decay to the lattice, as a function of $k^{2}/k_{M}^{2}$, for $\xi=0.1, 0.3, 0.9$.   The upper values are $\gamma^{+}$ and the lower values are $\gamma^{-}$.\\

\subsubsection{Significant Lattice Decay}
With lattice decay included, the results are a simple quadratic with complicated coefficients.  To display it, we employ the rates
\bea
w_{m}=\frac{1}{\tau_{mL}}, \quad w_{M}=\frac{1}{\tau_{ML}},
\label{defs2}
\eea
where $w_{m},w_{M}\rightarrow0$ as the lattice decay rates go to zero.  Then
\begin{widetext}
\begin{equation}
\gamma=\frac{r}{2}\Bigg((\frac{w_m}{r}+\frac{w_M}{r}+\frac{ k^{2} }{ k_{M}^{2} }+1)
\pm\Big[(\frac{w_m}{r}-\frac{w_M}{r}+\frac{ k^{2} }{ k_{M}^{2} }+1)^{2} - 4\xi(1-\frac{w_M}{r})(\frac{k^2}{k_M^2}+\frac{w_m}{r})\Big]^{1/2} \Bigg).\label{gammaL}
\end{equation}
\end{widetext}
Note that as $w_{m},w_{M}\rightarrow0$, \eqref{gammaL} goes to \eqref{gammaNoL}.

\begin{figure}[!htb]
\scalebox{0.35}[0.35]{\includegraphics{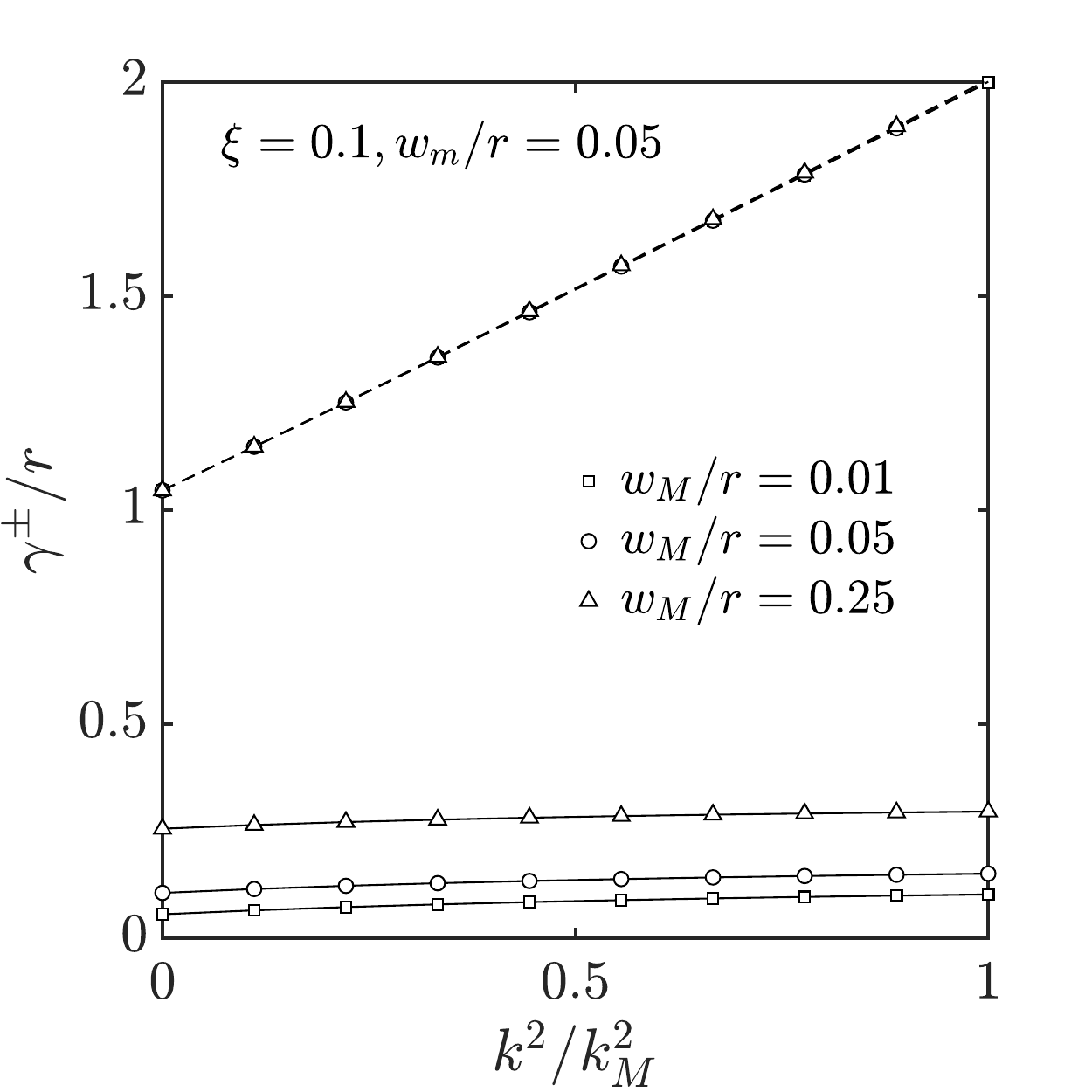}}
\caption{Eigenrates $\gamma$ relative to exchange-driven rate $r$, versus wavevector ratio $(k/k_{M})^{2}$, for three $M\rightarrow L$ lattice decay rates $w_{M}$ relative to $r$.  We take susceptibility ratio $\xi=0.01$ and $m\rightarrow L$ lattice decay rate $w_{m}$ relative to $r$ of $w_{m}/r=0.05$.  For rate $r$ and $\xi$ see \eqref{xi}; for wavevector $k_{M}$ see \eqref{kL2}.  
The dashed lines and the solid lines represent $\gamma^+$ and $\gamma^-$, respectively. }
\label{fig:modes_lattice_xi0.1}
\end{figure}

\begin{figure}[!htb]
\scalebox{0.35}[0.35]{\includegraphics{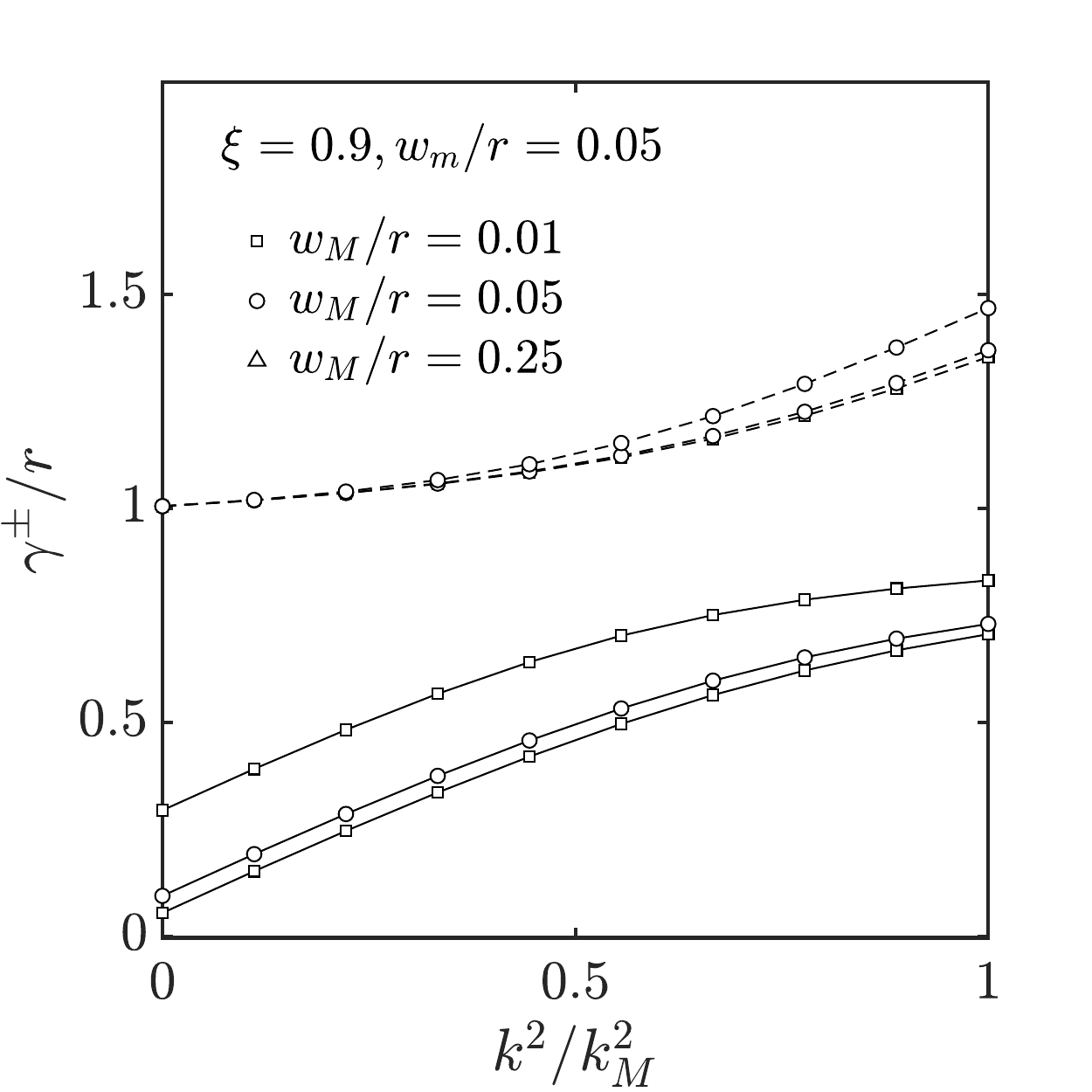}}
\caption{Eigenrates $\gamma$ relative to exchange-driven rate $r$, versus wavevector ratio $(k/k_{M})^{2}$, for three $M\rightarrow L$ lattice decay rates $w_{M}$ relative to $r$.  We take susceptibility ratio $\xi=0.09$ and $m\rightarrow L$ lattice decay rate $w_{m}$ relative to $r$ of $w_{m}/r=0.05$.   For rate $r$ and $\xi$ see \eqref{xi}; for wavevector $k_{M}$ see \eqref{kL2}.  
The dashed lines and the solid lines represent $\gamma^+$ and $\gamma^-$, respectively.}
\label{fig:modes_lattice_xi0.9}
\end{figure}

We believe it is a realistic simplification is to take the lattice decay rate relatively small compared to the exchange driven cross-decay between $M$ and $m$.  Using this approximation and taking 
$\tau_{mL}/r=0.05$, we present results for $\xi=0.1$ in Fig.~\ref{fig:modes_lattice_xi0.1} and for  $\xi=0.9$ in Fig.~\ref{fig:modes_lattice_xi0.9}.  The curve for $\xi=0.3$ (not shown) is very similar to that for $\xi=0.1$.  When lattice decay is included the lower mode for small $k$ is no longer purely diffusive.

For comparison we recall the standard diffusion equation with a decay term of a single degree of freedom (normally taken to be $M$).  With relaxation time $\tau$ and diffusion constant $D$, it is given by
\bea
\partial_t M =D \partial_x^2 M - \frac{1}{\tau}(M-M_{eq}),
\label{dM/dtMalone}
\eea
where $M_{eq}=M_{0}+\chi_{M}H$ can be applied to a paramagnet on taking $M_{0}=0$. 
The single decay rate $\gamma$ for the single variable $M$ is
\bea
\gamma=\frac{1}{\tau}+Dk^{2}.
\label{model_m_only}
\eea
This is to be contrasted with the predicted two decay rate behavior for the two variables $M$ and $m$.

For the steady-state problem, we may use this equation with $\tau\rightarrow\infty$, to obtain decay along $x$ of the form $e^{\pm k_{d}x}$ with inverse decay length, or decay wavevector,
\bea
k_{d}=(D\tau)^{-1/2}.
\label{k_d}
\eea
Although derived for a ferromagnet, it also applies to a paramagnet; we will use this result later. 

\subsection{Spacial Response to Frequency $\omega$}
Now consider that the system is subject to oscillation at a known real frequency $\omega$, so $dM,dm \sim e^{-i\omega t}$.  This can be done,  e.g., by injecting an ac spin current into the system.  The spatial response of $dM$ and $dm$ can be obtained by inverting the dispersion relation with $\gamma$ replaced by $i\omega$.  In general there will be a pair of complex values 
$k^+ = -k^-$, with a simple dependence on $\omega$ but with complicated coefficients.

\subsubsection{Slow Lattice Decay}
For $\tau_{ML},\tau_{mL}\rightarrow\infty$ we get
\bea
\frac{k^2}{k_{M}^{2}} = i\frac{\omega}{r} \frac{\omega + i r}{\omega+i\xi r}.
\eea
This is a pair of complex values $\pm k$, one exponentially growing and one exponentially decaying, with associated oscillations.

For $\omega\ll \xi r$ we have
\bea
\frac{k^{2}}{k_{M}^{2}}\rightarrow \frac{i\omega}{\xi r},
\label{klowwll}
\eea
and for $\omega\gg r$ we have
\bea
\frac{k^{2}}{k_{M}^{2}}\rightarrow \frac{i\omega}{r},
\label{klowwgg}
\eea

Once the eigenvalues are found, the eigenmodes, which give the relative amounts of $dm$ and $dM$, can be determined.  As usual, the physics is in the eigenmodes.

\subsubsection{Significant Lattice Decay}

\begin{figure}[!htb]
\scalebox{0.35}[0.35]{\includegraphics{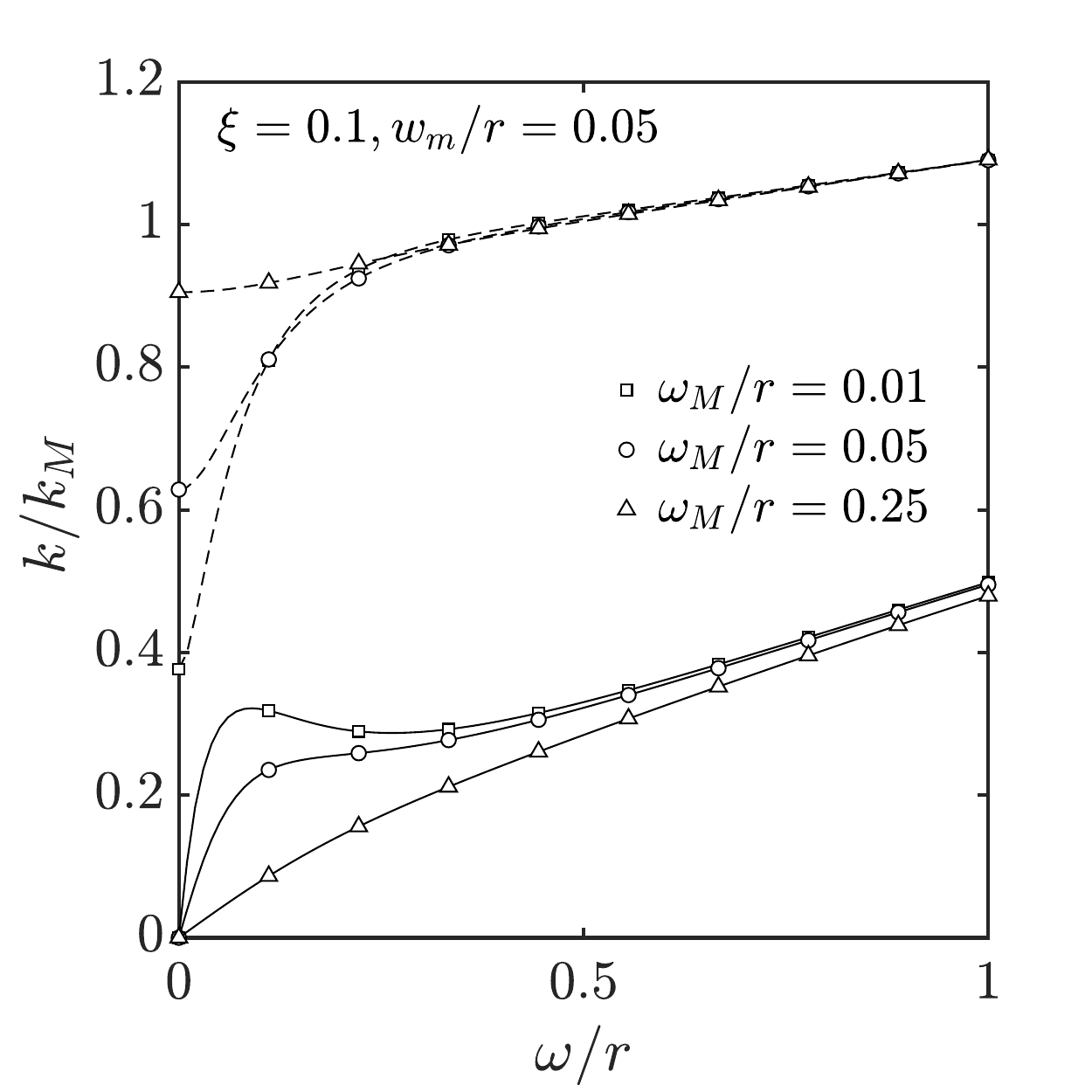}}
\caption{Wavevector $k$: real part is solid line and imaginary part is dashed line.   We take $m\rightarrow L$ lattice decay rate $w_{m}$ relative to $r$ of $w_{m}/r=0.05$.  We take three $M\rightarrow L$ lattice decay rates relative to $r$, or $w_{M}/r$. }
\label{fig: k_modes_lattice_xi0.1}
\end{figure}

For completeness we present $k^{2}$ when lattice decay is included, where $w_{m}$ and $w_{M}$ are defined in \eqref{defs2}.   Other than the dimensionless $\xi$, all symbols are rates:
\begin{equation}
 \frac{k^2}{k_M^2}=-\frac{w_m}{r}+\frac{(i\omega-r)(i\omega-w_M)}{r(i\omega+\xi w_M-w_M-\xi r)}.
\label{kwithLdecay}
\end{equation}
Fig.~\ref{fig: k_modes_lattice_xi0.1} presents the wavevector $k$; the solid line is the real part (oscillation) and the dashed line is the imaginary part (decay).

As $\omega\rightarrow0$, by working directly with the equations of motion we find that $k^{2}=-k_{F}^{2}$, so $k=\pm ik_{F}$, where
\begin{equation}
k_{F}^{2}=\frac{\tau_{M}}{D}\Big[\frac{1}{\tau_{Mm}\tau_{mL}}+\frac{1}{\tau_{mM}\tau_{ML}}+\frac{1}{\tau_{mL}\tau_{ML}}\Big].
\label{kF}
\end{equation}
All of three of these terms involve decay to the lattice.  We will now employ \eqref{kF} to study dc spin flux across a surface.

\section{dc Spin Flux Across a Surface}
\label{sect:boundaryconditions}
Consider a known rightward steady (dc) spin current from a paramagnet on the left ($x<0$) to a ferromagnet on the right ($x>0$), each treated as semi-infinite.  Within the paramagnet let there be a planar source at $x=-L$, where $L\gg k_{d}^{-1}$.  Let the source produce right and left spin currents with equal amplitudes, which decay on moving further further from $x=-L$.  We wish to find the reflected spin current in the paramagnet and the transmitted spin current in the ferromagnet. \\
\indent{}
{\bf Paramagnet:} We give the paramagnet $m_{p}$ an incoming rightward decaying mode of unit amplitude and decay wavevector $k=k_{d}$ given by \eqref{k_d}, and a leftward decaying reflected mode of unknown amplitude $A_{p}$:
\begin{equation}
dm_{p}=e^{-k_{d}x}+A_{p}e^{k_{d}x}. \quad (-L<x<0).
\label{dmPtrial}
\end{equation}
where, by \eqref{m_L} for $M=M_{0}=0$ in the paramagnet,
\begin{equation}
\delta m_{p}=dm_{p}=-\chi_{p}h_{p}^{*}.
\label{deltamp}
\end{equation}
The spin current then is given by
\begin{equation}
j_{pi}=-D\partial_{i}\delta m_{p}=D_{p}k_{d}(e^{-k_{d}x}-A_{p}e^{k_{d}x}).
\label{j_pdc}
\end{equation}
At $x=0^{-}$ the paramagnet has
\begin{equation}
j_{pi}^{0-}=D_{p}k_{d}(1-A_{p}).
\label{jmPsurface}
\end{equation}
\indent{} {\bf Ferromagnet:} We give the ferromagnet $m_{F}$ a decaying rightward mode of unknown amplitude $A_{F}$ and wavevector $k_{F}$ given by \eqref{kF}:
\begin{equation}
dm_{F}=A_{F}e^{-k_{F}x}.  \quad (x>0)
\label{dmFPtrial}
\end{equation}
\indent{}
In the steady-state (32) gives
\begin{equation}
dM=(-1+r\tau_{ML})dm_{F}.
\label{dM/dm}
\end{equation}
Then with $m\rightarrow m_{F}$, \eqref{delm_L} gives
\begin{eqnarray}
\delta m_{F}&=&dm_{F}+\xi dM =dm_{F}[1+\xi(-1+r\tau_{ML})^{-1})] \cr
&\equiv&(1+S)dm_{F}.
\label{deltamdc}
\end{eqnarray}
\indent{}
The spin current then is given by
\begin{equation}
j_{mi}=-D_{m}\partial_{i}\delta m_{F}=-D_{m}(1+S)\partial_{i}(dm_{F}),
\label{j_mdc}
\end{equation}
where, by \eqref{m_L}, with appended subscripts $F$, $\delta m_{F}=-\chi_{m}h^{*}_{F}$.  \\
\indent{}
Then, by \eqref{deltamdc},
\begin{equation}
\delta m_{F}=(1+S)A_{F}e^{-k_{F}x},
\label{deltamFtrial}
\end{equation}
so at $x=0^{+}$ for the ferromagnet \eqref{j_mdc} gives
\begin{equation}
j_{mi}^{0+}=D_{m}(1+S)k_{F}A_{F}.
\label{jmFsurface}
\end{equation}
\indent{}
{\bf Constraints:} With the two unknowns $(A_{p}, A_{F})$, there must be two constraints.  \\
\indent{}A first constraint comes from matching the spin currents at $x=0^{-}$ and $x=0^{+}$.  Thus
$j_{mi}^{0-}=j_{mi}^{0+}$, which leads to
\begin{equation}
D_{p}k_{d}(1-A_{p})=D_{m}(1+S)k_{F}A_{F}.
\label{matchjs}
\end{equation}
\indent{}A second constraint arises from the spin current crossing the interface being driven, in linear response, by the difference across the interface in the magnetoelectrochemical ``fields'' $h^{*}$ that act on $m$.\cite{JohnsonSilsbee87}  
We write the coefficient of linear response for spin diffusion ${\cal D}$.
Thus we take
\begin{equation}
j_{mi}={\cal D}\Delta h^{*}={\cal D}(h^{*0-}-h^{*0+}).
\label{jSi}
\end{equation}
${\cal D}$ has units of spin diffusivity divided by length, which is a velocity.
See Ref.~\onlinecite{JanossyMonod73,MenardWalker74,Flesner76}.  Perhaps the first time a surface transport velocity appeared was in Shockley's recombination velocity.\cite{Shockley50,KrcmarSaslow18}\\
\indent{}
On substitution, \eqref{jSi} explicitly leads to
\begin{equation}
D_{p}k_{d}(1-A_{p})=-{{\cal D}}\Big[\frac{1}{\chi_{p}}(1+A_{p})-\frac{1}{\chi_{m}}(1+S)A_{F}\big].
\label{jS2}
\end{equation}
\indent{}
In solving \eqref{matchjs} and \eqref{jS2} for $A_{p}$ and $A_{F}$ it is useful to define
\begin{equation}
C=\big(1-\frac{{\cal D}/\chi_{m}}{D_{m}k_{F}}\big)D_{p}k_{d}.
\label{C}
\end{equation}
Here $(D_{m}, \chi_{m}, k_{F})$ and $(D_{p}, \chi_{p}, k_{d})$ respectively refer to spin diffusion in the ferromagnet and in the paramagnet.  In terms of $C$ we have
\begin{equation}
A_{p}=\frac{C+{\cal D}/\chi_{p}}{C-{\cal D}/\chi_{p}}, 
\label{Ap}
\end{equation}
\begin{equation}
A_{F}=-\frac{D_{p}k_{d}}{D_{m}k_{F}(1+S)}\frac{2{\cal D}/\chi_{p}}{C-{\cal D}/\chi_{p}}. 
\label{AF}
\end{equation}
\indent{}
More complex situations can be treated using the above approach.  

\section{Summary}
\label{sect:Summary}

We have have developed the idea that a ferromagnet has two macroscopic longitudinal variables: the usual magnetization $M$ --- due to a statistical equilibrium distribution that cannot diffuse; 
and the spin accumulation $m$ --- due to a statistical non-equilibrium distribution that can diffuse.

By requiring that the phenomenological energy density be minimized for $M_{eq}=M_{0}+\chi_{M}H$ and $m_{eq}=0$, we find a new phenomenological exchange term between $M$ and $m$. It takes the form $-\lambda_{M}(M-M_{0})m$, with $\lambda_{M}=-\chi_{M}^{-1}$. The new statistical exchange term is not the thermally averaged exchange field, but rather what would arise from computing the free energy while requiring that $m=0$ in equilibrium.


Using the methods of Onsager's irreversible thermodynamics, we have found the equations of motion for $M$ and $m$, and have related the spin current to gradients of the deviation from local equilibrium $\delta m$.

We then studied the time decay for this system subject to an imposed real wavevector $k$, finding two rather than one decay mode; and the spatial decay of such a system subject to an imposed real frequency $\omega$, finding a single pair of modes (as for a single one magnetic variable), but with a more complicated frequency dependence than for simple diffusion.  The single degree of freedom theory has only one time-decay mode $\gamma$, and a pair of space-decay modes with a simpler dependence on frequency.


It would be of great value to have experimental studies of these predictions, which differ from what is predicted for a single ferromagnetic variable $M$ that is both non-zero in equilibrium and is responsible for spin diffusion. \\

\section{Implications}
\label{sect:Implications}
Micromagnetics normally deals with the response of the magnetization $\vec{M}$ normal to the equilibrium direction $\hat{M}_{eq}$, which can vary in space.  In the presence of $\vec{m}$ one can also develop corresponding coupled micromagnetics equations.\cite{Saslow17,SunSaslow19}  The present work shows how to include the longitudinal components.  Note that quantum-mechanical calculations include neither decay nor diffusion, nor do site-by-site studies of spin dynamics. 

Other ordered systems may have this property that an equilibrium statistical order parameter cannot diffuse, but that a non-equilibrium ``accumulation'' with the same symmetry can diffuse, with cross-decay between the two.  For example, a superconductor has a non-zero pair order parameter $\Delta$; as with $M$, perhaps $\Delta$ can decay but cannot diffuse; but the non-equilibrium pair order parameter ``accumulation'' $\delta$ can both decay and diffuse.  Of course there are strong non-dissipative restoring forces acting when $\Delta$ is out of equilibrium, and these will tend to mask the effect analogous to what we have studied for $M$ and $m$.

In closing we note the following.  One might think that number density $n$ might have similar properties to $M$ in a system where $n$ is non-uniform in equilibrium.  However, $n$ is a strictly conserved quantity, unlike $M$ and $\Delta$, which are statistically determined quantities depending for their existence upon a system that has condensed into an ordered state.  The hypothesis that is the basis of the present work depends crucially on the statistical nature of $M$.

\acknowledgements{}
We thank the organizers of the 7th Front Range Advanced Magnetics Symposium for the opportunity to present an early version of this work. C. S. is supported by the Fundamental Research Funds for the Central Universities from China.


\begin{thebibliography}{}
\bibitem{size} As the size of the system decreases, the possibility of diffusion increases.  We nevertheless think that even for nanometer sized systems, where other and unforeseen statistical effects may become important, diffusion of $M$ is unlikely.
\end{thebibliography}
\end{document}